\documentclass[10pt,twocolumn]{article}

\usepackage[utf8]{inputenc}
\usepackage[T1]{fontenc}
\usepackage[margin=0.75in]{geometry}
\usepackage{amsmath, amssymb, amsfonts, bm}
\usepackage{mathtools}

\usepackage{graphicx}
\usepackage{float}
\usepackage[caption=false, font=footnotesize]{subfig}
\usepackage{pgfplots}
\pgfplotsset{compat=1.18}

\usepackage[table]{xcolor}
\usepackage{cite}
\usepackage{hyperref}
\usepackage{url}
\hypersetup{
    colorlinks=true,
    linkcolor=blue!60!black,
    citecolor=blue!60!black,
    urlcolor=blue!60!black
}
\usepackage{booktabs}
\usepackage{multirow}
\usepackage{array}
\usepackage{tabularx}

\usepackage{enumitem}
\setlist[itemize]{leftmargin=*}
\setlist[enumerate]{leftmargin=*}

\usepackage{algorithm}
\usepackage{algpseudocode}
\floatname{algorithm}{Algorithm}

\algnewcommand{\Initialize}[1]{\State \textbf{Initialize:} #1}

\def\BibTeX{{\rm B\kern-.05em{\sc i\kern-.025em b}\kern-.08em
    T\kern-.1667em\lower.7ex\hbox{E}\kern-.125emX}}

\newcommand{\Norm}{\mathrm{Norm}}
\newcommand{\Enc}{\mathrm{Enc}}

\newcommand{\sigmaf}{\mathrm{sim}}

\newcolumntype{L}[1]{>{\raggedright\arraybackslash}p{#1}}
\newcolumntype{C}[1]{>{\centering\arraybackslash}p{#1}}
\newcolumntype{R}[1]{>{\raggedleft\arraybackslash}p{#1}}

\newcommand{\Description}[1]{}

\begin{document}

\title{MURAL: Multimodal Uncertainty-aware Recommendation via Adaptive edge Learning}

\author{%
\begin{minipage}{\dimexpr\textwidth-2\tabcolsep\relax}
\centering
\begin{minipage}[t]{0.32\textwidth}\centering
\textbf{Ahmad Mousavi}\textsuperscript{*}\\[2pt]
{\small Department of Mathematics and Statistics,\\ American University}\\
{\small Washington, DC, USA}\\
{\small\texttt{mousavi@american.edu}}
\end{minipage}\hfill
\begin{minipage}[t]{0.32\textwidth}\centering
\textbf{Majid Alikhani}\textsuperscript{*}\\[2pt]
{\small Independent Researcher}\\
{\small Toronto, Ontario, Canada}\\
{\small\texttt{alikhanimjd@gmail.com}}
\end{minipage}\hfill
\begin{minipage}[t]{0.32\textwidth}\centering
\textbf{Yeon-Chang Lee}\\[2pt]
{\small Department of Computer Science\\ and Engineering,\\ Ulsan National Institute of\\ Science and Technology}\\
{\small Ulsan, South Korea}\\
{\small\texttt{yeonchang@unist.ac.kr}}
\end{minipage}\\[1.4em]
\begin{minipage}[t]{0.32\textwidth}\centering
\textbf{Roberto Corizzo}\\[2pt]
{\small Department of Computer Science,\\ American University}\\
{\small Washington, DC, USA}\\
{\small\texttt{rcorizzo@american.edu}}
\end{minipage}\hfill
\begin{minipage}[t]{0.32\textwidth}\centering
\textbf{Yeganeh Abdollahinejad}\\[2pt]
{\small Department of Biosystems and\\ Agricultural Engineering,\\ Michigan State University}\\
{\small East Lansing, MI, USA}\\
{\small\texttt{yegi@msu.edu}}
\end{minipage}\hfill
\begin{minipage}[t]{0.32\textwidth}\mbox{}\end{minipage}
\end{minipage}%
}
\date{}

\twocolumn[
\begin{@twocolumnfalse}
\maketitle
\begin{center}
\begin{minipage}{0.86\textwidth}
\begin{center}\large\textbf{Abstract}\end{center}
\noindent
Multimodal Graph Neural Networks have become standard for recommendation by augmenting sparse interaction data with content features. Yet current architectures face two bottlenecks: structural rigidity, from a reliance on static precomputed similarity graphs that cannot adapt to evolving preferences; and semantic fragility, where noisy modality signals are indiscriminately fused, distorting the collaborative signal. We propose MURAL (\textbf{M}ultimodal \textbf{U}ncertainty-aware \textbf{R}ecommendation via \textbf{A}daptive edge \textbf{L}earning), a unified framework that shifts multimodal recommendation from fixed structural augmentation to dynamic topology discovery. To address structural rigidity, an Adaptive Edge Learner combines a differentiable retrieval-augmented strategy with an approximate nearest neighbor search to discover latent item-item correlations that are both semantically adaptive and computationally scalable ($O(N \log N)$). To address semantic fragility, an Uncertainty-Aware Fusion module models the aleatoric uncertainty of heterogeneous modalities, dynamically down-weighting unreliable features while prioritizing high-confidence signals as a defense against cross-modal noise. We further employ a contrastive teacher-student alignment that anchors modality-specific representations to stable behavioral signals, ensuring optimization stability without gradient leakage. Experiments on large-scale benchmarks including TikTok and Amazon show that MURAL significantly surpasses both structural and generative state-of-the-art baselines, achieving superior accuracy while offering interpretability through domain-specific modality dominance and robustness under extreme data corruption.

\vspace{0.9em}
\noindent\textbf{Keywords:} Graph Neural Network, Multimedia Recommendation, Representation Learning
\end{minipage}
\end{center}
\vspace{2.2em}
\end{@twocolumnfalse}
]

\begingroup
\renewcommand{\thefootnote}{\ensuremath{*}}%
\footnotetext{Ahmad Mousavi and Majid Alikhani contributed equally to this work.}%
\endgroup

\section{Introduction}
Multimedia recommendation systems have become indispensable components of modern web-based applications, ranging from large-scale e-commerce platforms to diverse content-sharing ecosystems \cite{alamdari2020systematic, deldjoo2020recommender}. Unlike traditional collaborative filtering, these systems leverage a rich array of item modalities - including textual descriptions, visual frames, and acoustic signals to capture fine-grained user preferences and item characteristics \cite{10.1145/3397271.3401132}. By integrating heterogeneous signals, multimedia recommenders have shown the capability to alleviate the fundamental challenges of data sparsity, and effectively construct robust latent representations even when historical user-item interactions are scarce \cite{raza2025comprehensivereviewrecommendersystems, natarajan2020resolving}.

The evolution of multimedia recommendation has progressed from early feature-concatenation methods like VBPR \cite{he2015vbprvisualbayesianpersonalized} to sophisticated attention-based architectures such as ACF \cite{10.1145/3077136.3080797}, which identify component-level preferences. Recently, GNNs have emerged as the state-of-the-art paradigm, with models such as MMGCN \cite{10.1145/3343031.3351034}, GRCN \cite{10.1145/3394171.3413556}, and LATTICE \cite{Zhang_2021} effectively leveraging graph structures to propagate multimodal information across user-item and item-item relations. By aggregating features from neighboring nodes, these models learn latent representations that capture both structural collaborative signals and raw content attributes. More recently, self-supervised learning (SSL) and generative models have been introduced to further enhance robustness. For instance, MMSSL utilizes adversarial training \cite{Wei_2023}, while DiffMM \cite{zhao2024denoisingdiffusionrecommendermodel} and DiffCL \cite{11060893} employ diffusion models to denoise embeddings. Moreover, advances such as AlignRec \cite{10.1145/3627673.3679626} and AB-Rec \cite{10.1145/3711896.3737275} have significantly improved performance by decomposing the recommendation objective into multi-stage alignment tasks and balancing optimization gradients. However, these methods often treat the underlying item-item topology as static and assume uniform modality reliability across items.

A critical challenge in Multimodal Graph Neural Networks (MGNNs) is the "Mirroring Effect": when modality-aware graphs are constructed based on shared interaction data, they tend to converge toward the topology of the original interaction graph \cite{sone2025mmgsl}. This structural redundancy prevents the model from discovering latent item-item correlations that exist purely within the multimodal space; signals that are essential for recommending items with sparse behavioral links.

While recent structural learners have attempted to address this by introducing explicit item-item edges \cite{wang2021dualgnn, tao2020mgat, Zhang_2021,Zhou_2023,sone2025mmgsl}, they remain constrained by two fundamental bottlenecks:
\begin{enumerate}
    \item Structural Rigidity: Most existing frameworks \cite{Zhou_2023,sone2025mmgsl}
    rely on static, pre-computed similarity heuristics 
    to define the graph topology. This assumes that the optimal semantic structure is fixed and known a priori, ignoring the fact that semantic relationships should evolve dynamically alongside the recommendation objective.
    \item Semantic Fragility: Current architectures \cite{Zhang_2021,Zhou_2023,sone2025mmgsl} often adopt a deterministic fusion strategy, where content and behavioral embeddings are combined without accounting for the inherent noise or reliability of heterogeneous modalities. In real-world environments like micro-videos or e-commerce, modality quality is highly inconsistent; treating a noisy visual feature with the same weight as a high-fidelity textual description can distort the collaborative filtering signal.
\end{enumerate}

In this paper, we propose MURAL (Multimodal Uncertainty-aware Recommendation via Adaptive edge Learning), a unified framework that moves beyond static structural enrichment toward dynamic topology discovery and robust fusion. The main contributions of this work are summarized as follows:
\begin{itemize}
    \item Adaptive Topology Discovery: We shift the paradigm from static, heuristic-based graph construction to a differentiable approach. By introducing an Adaptive Edge Learner (AEL), MURAL treats graph structure as a learnable task, discovering evolving semantic correlations that standard similarity metrics overlook while maintaining $O(N \log N)$ scalability through retrieval-augmented search.
    \item Aleatoric Uncertainty-Aware Fusion: We address the inherent "semantic fragility" of heterogeneous data by explicitly modeling modality reliability. Our UAF mechanism learns to quantify aleatoric uncertainty, dynamically down-weighting noisy or sparse signals (e.g., generic descriptions or trending background audio) to ensure the final representation is anchored by the most discriminative features.
    \item Stabilized Behavioral-Semantic Anchoring: To resolve the optimization instabilities and representation collapse common in current alignment frameworks, we propose a simple gradient-detached anchoring strategy. This design decouples structural discovery from feature alignment, providing a scalable and robust alternative to complex cross-modality encoders.
    \item Unified Graph Propagation Framework: We integrate these modules into MURAL, a unified framework that consistently outperforms structural, generative, and alignment-based state-of-the-art baselines across multiple large-scale benchmarks.
\end{itemize}


\section{Related Work}
The integration of multimodal content has become essential in modern recommender systems \cite{liu2022disentangledmultimodalrepresentationlearning, xu2025surveymultimodalrecommendersystems}, particularly for mitigating data sparsity and enriching item representation \cite{wei2021contrastivelearningcoldstartrecommendation, 10.1145/3640457.3688009}. The shift from interaction-only models to multimodal recommender systems (MMRS) was pioneered by early hybrid models like VBPR \cite{he2015vbprvisualbayesianpersonalized}, which integrated visual features into the Bayesian Personalized Ranking framework. Later on, methods like CKE \cite{10.1145/2939672.2939673}, and JRL \cite{10.1145/3132847.3132892} combined textual, visual, and structural features to further improve recommendation quality. With the rise of Graph Neural Networks (GNNs), graph-based approaches emerged that leverage high-order neighborhood information to enhance user and item representations \cite{Wu_2021, 10.1145/3535101}. The foundation of modern graph-based recommendation lies in LightGCN \cite{he2020lightgcnsimplifyingpoweringgraph}, which simplified the GCN architecture \cite{kipf2017semisupervisedclassificationgraphconvolutional} by focusing exclusively on neighborhood aggregation to capture collaborative signals. To incorporate auxiliary content, MMGCN \cite{10.1145/3343031.3351034} pioneered the use of modality-specific message-passing channels, allowing visual and textual features to propagate through independent graph structures. Subsequent works like GRCN \cite{10.1145/3394171.3413556} introduced adaptive reweighting to prune noisy interaction edges. However, these methods are primarily "interaction-centric," meaning they are fundamentally limited by the observed user-item bipartite graph. In highly sparse interaction scenarios, this reliance prevents the model from capturing deep semantic associations that exist independently of user behavior.

To mitigate interaction sparsity, Graph Structure Learning (GSL) has emerged to uncover latent item topologies. LATTICE and MMGSL \cite{sone2025mmgsl} pioneered this approach by mining graphs from pre-computed feature similarities. However, these approaches are often constrained by heuristic neighborhood initialization. 
This topological bias prevents the model from discovering latent semantic connections that only emerge through behavioral signals. Furthermore, these methods typically assume uniform reliability across modalities, failing to account for item-level content noise that can distort the structural signal during message passing.

\begin{figure*}[t]
    \centering
    \includegraphics[height=6cm, width=1\textwidth]{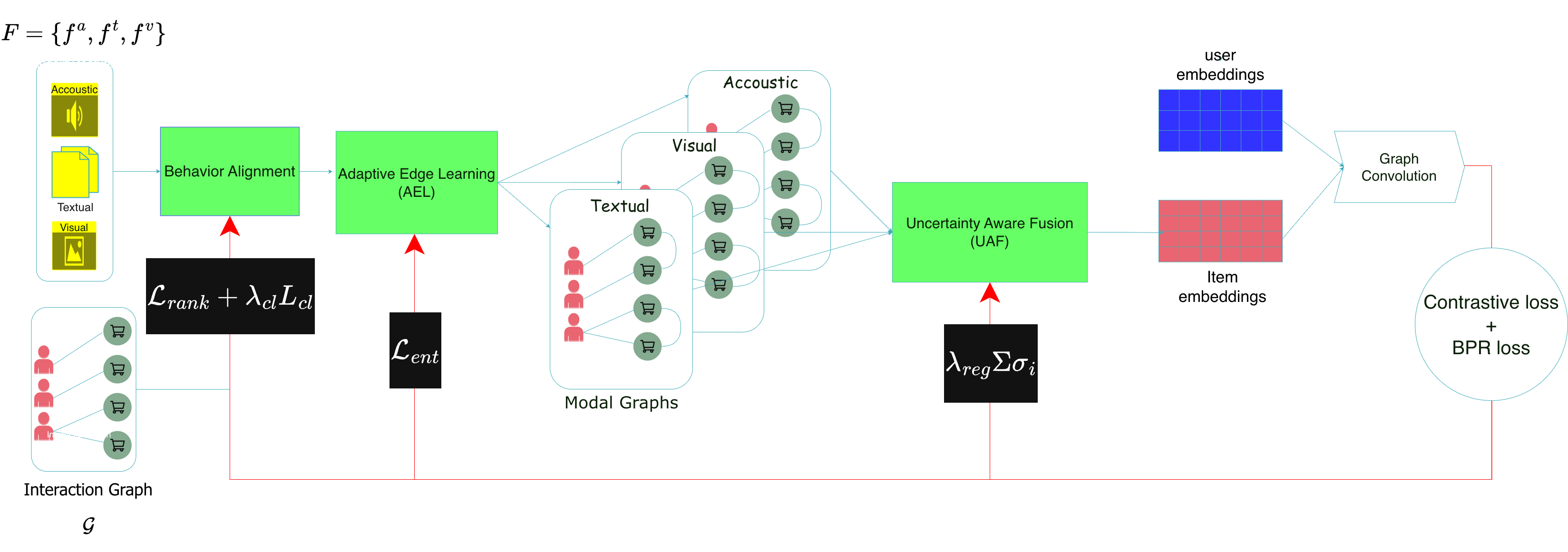}
    \caption{Overview of MURAL}
    \Description{A diagram illustrating the MURAL framework, including the adaptive edge learner, modality-specific graph construction, contrastive alignment module, and uncertainty-aware fusion component.}
    \label{fig:AEL-UAF}
\end{figure*}

Self-Supervised Learning (SSL) has emerged as a powerful paradigm for aligning multimodal features with collaborative signals \cite{jiang2023adaptivegraphcontrastivelearning, Lin_2022, liu2022unsuperviseddeepgraphstructure}. Foundational works such as SGL \cite{Wu_2021} and NCL \cite{Lin_2022} introduced structural and neighborhood-enriched augmentations to maximize agreement between different graph views. Building on these, SimGCL \cite{zhao2025moresimplegraphcontrastive} demonstrated that simpler, non-structural augmentations---such as injecting uniform noise into the embedding space---can achieve superior performance by avoiding the bias of random edge dropping. Building on these structural advances, multimodal SSL frameworks like MMSSL \cite{Wei_2023} and AlignRec \cite{10.1145/3627673.3679626} employ contrastive objectives (e.g., InfoNCE) to enforce consistency across different modality views. More recently, generative approaches like DiffMM \cite{jiang2024diffmmmultimodaldiffusionmodel} and DiffCL \cite{11060893} have utilized diffusion-based denoising to refine node representations.

While current multimodal SSL frameworks have made significant strides, they are collectively hindered by a dual dependency on structural rigidity and semantic fragility. Specifically, these methods rely on static, pre-computed item topologies and apply an indiscriminate alignment pressure, ignoring the inherent aleatoric uncertainty across diverse modalities. MURAL breaks this paradigm by transitioning from fixed heuristics to dynamic topology discovery via its Adaptive Edge Learner, while simultaneously employing uncertainty-aware fusion to ensure that contrastive signals are filtered and weighted based on their objective reliability.

\section{Adaptive Edge Learning with Uncertainty-Aware Fusion}
We propose MURAL, a unified framework that transitions multimodal recommendation from static structural augmentation to dynamic, reliability-aware discovery. Unlike prior models constrained by pre-computed heuristics and deterministic weighting, MURAL introduces a differentiable architecture to learnably determine: (i) the latent topology of item-item semantic correlations, (ii) the adaptive dominance of heterogeneous modalities during fusion, and (iii) the aleatoric uncertainty of specific signals to ensure noise-robust alignment.
Fig.~\ref{fig:AEL-UAF} shows an overview of the proposed framework.

\paragraph{Problem Setup}
Let $\mathcal U=\{u_1,\dots,u_{N_u}\}$ and $\mathcal I=\{i_1,\dots,i_{N_i}\}$ denote users and items, respectively.
We observe a binary interaction matrix
$R\in\{0,1\}^{N_u\times N_i}$,
where $R_{ui}=1$ if user $u$ has interacted with item $i$ and $0$ otherwise.
Each item $i$ has multimodal content features 
$\mathcal F_i=\{f_i^{(1)},f_i^{(2)},\dots,f_i^{(M)}\}$,
corresponding to $M$ modalities (e.g., text, image, and audio).
Our goal is to learn low-dimensional embeddings 
$h_u,h_i\in\mathbb R^d$
for each user and item that accurately predict interaction likelihood,
\[
\hat y_{ui}= \langle h_u, h_i\rangle,
\]
while efficiently incorporating multimodal information.

\paragraph{Modality Encoders} 
For each modality $m \in \{1,\dots,M\}$, 
we obtain a modality-specific embedding from either raw content or pretrained representations. 
Formally, we define
\begin{equation}
e_i^{(m)} = \Norm\big(W_m\, \Enc^{(m)}(f_i^{(m)})\big)
\label{eq:modprojection}
\end{equation}
where $\Enc^{(m)}(\cdot)$ denotes a modality encoder (e.g., BERT for text, ViT for images, AudioLM for audio) 
and $W_m \in \mathbb{R}^{d \times d_m}$ is a learnable linear projection that maps the encoder output 
from dimension $d_m$ into a shared latent dimension $d$.
This projection aligns heterogeneous modality spaces and ensures all $e_i^{(m)}$ reside in a common space for fusion and contrastive alignment.
When pretrained modality embeddings are already available (e.g., CLIP features), 
$\Enc^{(m)}$ can be treated as the identity and $W_m$ may be initialized as the identity or omitted.
All embeddings are $\ell_2$-normalized through $\Norm(\cdot)$ for stability, 
and encoders may remain frozen or be lightly adapted using low-rank (LoRA) updates to preserve efficiency.

\subsection{Adaptive Graph Construction}
\paragraph{Behavior-Aligned Modality Representations}
To make each modality embedding consistent with user behavior, we construct a \emph{behavior-aligned} representation that blends each modality's content signal with the collaborative signal learned from user-item interactions.

Let $h_i \in \mathbb{R}^d$ denote the current \emph{behavioral embedding} of item $i$, 
obtained from the interaction graph through the propagation defined in Eq.~(\ref{eq:prop}). 
Intuitively, $h_i$ encodes how users collectively perceive and interact with item $i$, 
capturing behavioral similarity (items co-consumed or co-rated by similar users) rather than raw content similarity. 
By contrast, $e_i^{(m)}$ captures modality-specific content features (e.g., textual, visual, or acoustic attributes) 
that may not always align with user preferences. 

To reconcile these two views, we define the behavior-aligned modality representation through a lightweight fusion:
\begin{equation}
\hat e_i^{(m)} = 
\kappa_0\,\operatorname{stopgrad}(h_i)
+ \kappa_1\, e_i^{(m)}, 
\qquad
\kappa_0+\kappa_1=1
\label{eq:aligned}
\end{equation}
where $\operatorname{stopgrad}(h_i)$ denotes a gradient-detached copy of $h_i$. 
This detachment prevents gradient leakage from the modality branches back into the interaction encoder, 
ensuring that the alignment process does not distort the underlying collaborative topology. 
The learnable coefficients $\kappa_0$ and $\kappa_1$ govern the balance between behavioral and content-based information. To prevent shortcut learning during optimization, we isolate $\kappa_0$ by parameterizing it as a learnable weight updated exclusively via the structural contrastive loss ($\mathcal{L}_{\mathrm{cl}}$), thereby shielding it from direct optimization by the main recommendation objective.
While prior works like AlignRec \cite{10.1145/3627673.3679626} employ heavy cross-attention mechanisms for alignment, MURAL shifts the objective toward adaptive graph structure learning and modality reliability estimation. We purposefully adopt a streamlined weighted-sum formulation. This ensures high throughput during the frequent graph rebuilds required by our Adaptive Edge Learner. Crucially, any noise or misalignment introduced by this simple formulation is explicitly handled by the subsequent Uncertainty-Aware Fusion (UAF) module, which calculates item-specific variances ($\sigma_i^{(m)}$) to down-weight unreliable signals that a fixed weighted sum would otherwise amplify.

\paragraph{Retrieval-Augmented Candidate Neighborhoods}
To maintain computational tractability on large-scale datasets, we avoid the $O(N_i^2)$ cost of exhaustive pairwise similarity by employing a two-stage retrieval-and-refine strategy. 

In the first stage, we construct a global Hierarchical Navigable Small World (HNSW) index \cite{malkov2018efficientrobustapproximatenearest}---a state-of-the-art ANN structure optimized for high-speed inner product search. The index is built over the fused item representations $\{\tilde h_i\}$, which encapsulate both behavioral signals and multimodal content. For every item $i$, we retrieve a candidate neighborhood $\mathcal N_r(i)$ of size $K_r$ ($K_r \ll N_i$), effectively pruning the search space.

To ensure the graph topology remains synchronized with the evolving latent space, we adopt a stochastic refresh schedule. Rather than rebuilding the index every epoch---which would introduce unnecessary overhead---we periodically re-index the item manifold throughout training. During the intervening epochs, the candidate sets $\mathcal N_r(i)$ remain fixed, while the Adaptive Edge Learner (see below) dynamically re-weights the edge strengths $s_{ij}^{(m)}$ based on the current parameters. This amortized approach maintains a complexity of $O(N_i \log N_i)$, making the framework viable for production-scale ecosystems like TikTok or Amazon. In our experiments, re-indexing every 10 epochs struck an optimal trade-off between index staleness and computational overhead.

\paragraph{Adaptive Edge Learning}
After retrieving the candidate set $\mathcal N_r(i)$ for each item $i$, the model learns modality-specific adjacency weights that capture how strongly each neighbor $j \in \mathcal N_r(i)$ 
is related to $i$ under modality $m$. 
Rather than using fixed similarities (e.g., cosine distance), we learn an adaptive scoring function parameterized by a small multilayer perceptron (MLP).

For each modality $m$, the scoring network 
$\phi_\theta^{(m)} : \mathbb{R}^{4d} \to \mathbb{R}$ 
takes as input a concatenation of behavioral and modality features for the item pair $(i, j)$:
\begin{equation}
s_{ij}^{(m)} =
\phi_\theta^{(m)}\!\big(
[\hat e_i^{(m)} \Vert \hat e_j^{(m)} 
 \Vert h_i \Vert h_j]
\big)
\label{eq:score}
\end{equation}
where $\Vert$ denotes concatenation and each component is $d$-dimensional. 
Thus, the input dimension to $\phi_\theta^{(m)}$ is $4d$. 
Including both $\hat e_i^{(m)}$ and $h_i$ allows the MLP to jointly consider modality-level and behavioral similarities 
when predicting the edge strength between two items.

Among the $K_r$ retrieved candidates, only the top-$k_m$ neighbors with the highest $s_{ij}^{(m)}$ values are retained to form a sparse, learnable adjacency matrix. 
The corresponding edge weights are normalized through a local softmax:
\begin{equation}
A_{ij}^{(m)} =
\frac{\exp(s_{ij}^{(m)})}
{\sum_{j'\in\mathcal N_k(i)} \exp(s_{ij'}^{(m)})},
\quad j \in \mathcal N_k(i)
\label{eq:adj}
\end{equation}
where $\mathcal N_k(i)\subseteq \mathcal N_r(i)$ contains the $k_m$ highest-scoring neighbors for item $i$. 
The resulting matrix $A^{(m)} \in \mathbb{R}^{N_i \times N_i}$ 
is sparse and modality-specific, encoding the relational structure discovered by the network rather than imposed by a fixed similarity metric.

To discourage overly sharp neighbor distributions, 
we apply an entropy regularizer
\[
\mathcal L_{\mathrm{ent}}
= \!\sum_{i,m}\sum_{j}A_{ij}^{(m)}\log A_{ij}^{(m)}
\]
weighted by a small coefficient $\lambda_{\mathrm{ent}}$. 
This term encourages diversity in the learned neighbors and prevents the adjacency matrix from collapsing into one-hot edges.

Overall, this two-stage process --- ANN retrieval followed by adaptive edge reweighting --- achieves both efficiency and flexibility: the coarse retrieval ensures scalability, while the learned scoring allows each modality to refine 
and personalize its graph topology based on both multimodal and behavioral context.

\subsection{Uncertainty-Aware Modality Fusion}
Not all modalities are equally informative. To address this, we move beyond deterministic fusion by explicitly modeling for each item $i$ the aleatoric uncertainty $\sigma_i^{(m)}>0$ associated with each signal along with an importance weight $a_i^{(m)}$ (logit) and its normalized attention $\alpha_i^{(m)}$:
\begin{align}
\alpha_i^{(m)} 
&= \frac{\exp(a_i^{(m)})}
        {\sum_{m'}\exp(a_i^{(m')})}, \\
\tilde h_i 
&= 
\sum_{m=1}^M
\alpha_i^{(m)}\,\frac{1}{(\sigma_i^{(m)})^2}\,e_i^{(m)}
\label{eq:fusion}
\end{align}
For numerical stability and to ensure that the variance remains strictly positive, we learn the log-variance $s_i^{(m)} = \log (\sigma_i^{(m)})^2$ as a per-item, per-modality parameter. These parameters are initialized to zero ($s_i^{(m)} = 0$), such that the initial weighting factor $\exp(-s_i^{(m)})$ is $1$ for all modalities.

The weights $\alpha_i^{(m)}$ are produced by a lightweight fusion network
$\operatorname{MLP}_\alpha([e_i^{(1)},\dots,e_i^{(M)},h_i]).$
A regularizer $\sum_{i,m}\log\sigma_i^{(m)}$ prevents the variances from collapsing to zero.

\subsection{Graph Propagation} We combine the interaction graph with all modality graphs through a weighted mixture. Let $\mathcal{M}$ be the set of modalities and $\hat{A}^{(UI)}\in\mathbb{R}^{(|U|+|I|)\times(|U|+|I|)}$ denote the normalized user--item interaction adjacency matrix, where each edge $(u,i)$ corresponds to a historical interaction $(R_{ui} = 1)$. Similarly, let $\hat{A}^{(m)}\in\mathbb{R}^{|I|\times|I|}$ ($m\in\mathcal{M}$) denote the row-normalized item--item adjacency matrix constructed for each modality through adaptive edge learning. Because $\hat{A}^{(m)}$ connects only items, it is expanded into a full $(|\mathcal{U}|+ \mathcal{|I|})\times(|\mathcal{U}|+|\mathcal{I}|)$ block matrix by placing it in the item--item sub-block, with zeros elsewhere:
\begin{equation}
\hat{A}^{(m)}_{full} = \begin{bmatrix}
0 & 0 \\
0 & \hat{A}^{(m)}
\end{bmatrix}
\end{equation}
We then form a convex combination of all adjacency matrices:
\begin{equation}
\hat A =
\lambda_{\mathrm{UI}}\hat A^{(\mathrm{UI})}
+ \sum_{m=1}^M \lambda_m \hat A^{(m)}_{full},
\qquad
\sum_m\lambda_m+\lambda_{\mathrm{UI}}=1
\label{eq:mix}
\end{equation}
where each $\hat A^{(\cdot)}$ is row-normalized and $\lambda_m$ and $\lambda_{\mathrm{UI}}$ are learnable nonnegative coefficients that balance the strength of behavioral
and modality-based connections.
We then propagate user and item embeddings with a LightGCN-style update:
\begin{equation}
H^{(\ell+1)}=\hat A\,H^{(\ell)},
\qquad
H^{(0)}=[H_U;H_I]
\label{eq:prop}
\end{equation}
After $L$ layers, the final embedding of each node $v$ is the mean of all layers:
\[
h_v=\frac{1}{L+1}\sum_{\ell=0}^L H_v^{(\ell)}
\]
Optionally, a skip connection combines propagation and fused embeddings:
$h_i\!\leftarrow\!\beta h_i + (1-\beta)\tilde h_i$.

\paragraph{Contrastive Alignment} To maintain consistency between behavior and content,
MURAL adds dual contrastive losses.

(a) Behavior--Modality alignment:
\begin{equation}
\mathcal L_{\mathrm{BM}}^{(m)} =
-\!\sum_i 
\log
\frac{
\exp(\sigmaf(h_i,e_i^{(m)})/\tau)
}{
\sum_j \exp(\sigmaf(h_i,e_j^{(m)})/\tau)
}
\label{eq:bm}
\end{equation}

(b) Cross-modality alignment:
\begin{equation}
\mathcal L_{\mathrm{MM}} =
-\!\sum_{m_x<m_y}\sum_i 
\log
\frac{
\exp(\sigmaf(e_i^{(m_x)},e_i^{(m_y)})/\tau)
}{
\sum_j \exp(\sigmaf(e_i^{(m_x)},e_j^{(m_y)})/\tau)
}
\label{eq:mm}
\end{equation}
Here $m_x,m_y\in\mathcal{M}$ and $\sigmaf(\cdot,\cdot)$ denotes cosine similarity and $\tau$ is the temperature parameter.

\paragraph{Preference Learning}
User--item affinity is predicted via the inner product
$\hat y_{ui}=\langle h_u,h_i\rangle$.
We adopt the Bayesian Personalized Ranking (BPR) loss:
\begin{equation}
\mathcal L_{\mathrm{rank}}
= -\!\!\!\sum_{(u,i^+,i^-)}
\!\!\!\log\sigma(\hat y_{ui^+}-\hat y_{ui^-})
\label{eq:bpr}
\end{equation}
where $i^+$ and $i^-$ are positive and negative samples for user $u$.

\paragraph{Overall Objective}
The full training loss combines ranking, contrastive, entropy, and regularization terms:

\begin{align}
\mathcal L &= 
\mathcal L_{\mathrm{rank}} 
+ \lambda_{\mathrm{cl}} \left( \sum_m \mathcal L_{\mathrm{BM}}^{(m)} + \mathcal L_{\mathrm{MM}} \right) 
+ \lambda_{\mathrm{ent}} \mathcal L_{\mathrm{ent}} \notag \\
&\quad + \lambda_{\mathrm{reg}} \left( \|\Theta\|_2^2 + \sum_{i,m} \log \sigma_i^{(m)} \right)
\label{eq:final}
\end{align}

where $\Theta$ denotes all learnable parameters.
Optimization is performed with the Adam optimizer.

\paragraph{Complexity and Scalability}
Let $N_i$ be the number of items, $K_r$ the candidate size, and $d$ the embedding dimension.
Edge scoring (\ref{eq:score}) costs $O(MN_iK_r d)$,
propagation (\ref{eq:prop}) costs $O((|\mathcal E_{\mathrm{UI}}|+N_i\sum_m k_m)d)$,
and ANN retrieval costs $O(N_i\log N_i)$ amortized.
Therefore, total complexity is linear in $N_i$ and $K_r$, scalable to large multimodal datasets.

\section{Evaluation}

\subsection{Experimental Setup}

\paragraph{Datasets}
To evaluate the effectiveness of MURAL, we conduct experiments on three widely used multimodal recommendation benchmarks: TikTok, Amazon-Baby, and Amazon-Sports. The TikTok dataset provides a dense multimodal representation of short-form video interactions, integrating visual, acoustic, and textual signals. Textual metadata, including captions and comments, was transformed into semantic vectors using Sentence-BERT. In the Amazon-Baby and Amazon-Sports categories, product titles and descriptions were encoded similarly via Sentence-BERT, while visual characteristics were captured through 4096-dimensional image embeddings.
The statistical details of the datasets are provided in Table \ref{tab:data_stats}.
\begin{table}[t]
\centering
\caption{Statistics of the datasets}
\label{tab:data_stats}
\setlength{\tabcolsep}{3pt}
\resizebox{\columnwidth}{!}{%
\begin{tabular}{l|ccc|cc|cc}
\toprule
\textbf{Dataset} & \multicolumn{3}{c|}{\textbf{TikTok}} & \multicolumn{2}{c|}{\textbf{Amazon-Baby}} & \multicolumn{2}{c}{\textbf{Amazon-Sports}} \\
\midrule
Modality & Visual & Audio & Text & Visual & Text & Visual & Text \\
Embed Dim & 128 & 128 & 768 & 4096 & 1024 & 4096 & 1024 \\
\midrule
\# Users & \multicolumn{3}{c|}{9,319} & \multicolumn{2}{c|}{19,445} & \multicolumn{2}{c}{35,598} \\
\# Items & \multicolumn{3}{c|}{6,710} & \multicolumn{2}{c|}{7,050} & \multicolumn{2}{c}{18,357} \\
\# Interactions & \multicolumn{3}{c|}{59,541} & \multicolumn{2}{c|}{139,110} & \multicolumn{2}{c}{256,308} \\
Sparsity & \multicolumn{3}{c|}{99.904\%} & \multicolumn{2}{c|}{99.899\%} & \multicolumn{2}{c}{99.961\%} \\
\bottomrule
\end{tabular}%
}
\end{table}

\paragraph{Evaluation Protocol}
To evaluate the performance of the top-$K$ recommendations, we utilize two widely recognized metrics: Recall@$K$ (R@$K$), and Normalized Discounted Cumulative Gain (NDCG@$K$). Crucially, to avoid the bias associated with sampled metrics---which can lead to artificially inflated results---we adopt the full-ranking strategy. For each user in the test set, we rank the ground-truth items against all items in the data set that the user has not interacted with.

\begin{table*}[t]
\centering
\caption{Performance comparison of MURAL and baseline models. The best results are highlighted in \textbf{bold}.}
\label{tab:main_results}
\small
\setlength{\tabcolsep}{4pt} 
\resizebox{\textwidth}{!}{%
\begin{tabular}{l|cc|cc|cc}
\toprule
\textbf{Dataset} & \multicolumn{2}{c|}{\textbf{TikTok}} & \multicolumn{2}{c|}{\textbf{Amazon-Baby}} & \multicolumn{2}{c}{\textbf{Amazon-Sports}} \\
\midrule
Metric & R@20 & N@20 & R@20 & N@20 & R@20 & N@20 \\
\midrule
LightGCN & 0.0730 & 0.0309 & 0.0649 & 0.0291 & 0.0746 & 0.0241 \\
MMGCN & 0.0664 & 0.0279 & 0.0714 & 0.0230 & 0.0703 & 0.0259 \\
LATTICE & 0.0855 & 0.0369 & 0.0843 & 0.0370 & 0.0921 & 0.0429 \\
DualGNN & 0.0834 & 0.0351 & 0.0829 & 0.0353 & 0.0881 & 0.0402 \\
MMSSL & 0.0900 & 0.0393 & 0.0944 & 0.0403 & 0.0980 & 0.0442 \\
DiffMM & 0.1102 & 0.0442 & 0.0954 & 0.0401 & 0.1002 & 0.0448 \\
FREEDOM & 0.1100 & 0.0458 & 0.0950 & 0.0403 & 0.1050 & 0.0485 \\
AlignRec & 0.1119 & 0.0473 & \underline{0.1008} & \underline{0.0441} & \underline{0.1127} & \underline{0.0501} \\
DiffCL & \underline{0.1158} & \underline{0.0490} & 0.0990 & 0.0421 & 0.1073 & 0.0498 \\
MMGSL & 0.1152 & 0.0488 & 0.0971 & 0.0413 & 0.1032 & 0.0464 \\
\midrule
\textbf{MURAL} & \textbf{0.1221}{\scriptsize$\pm0.0029$} & \textbf{0.0541}{\scriptsize$\pm0.0023$} & \textbf{0.1068}{\scriptsize$\pm0.0019$} & \textbf{0.0502}{\scriptsize$\pm0.0020$} & \textbf{0.1173}{\scriptsize$\pm0.0017$} & \textbf{0.0543}{\scriptsize$\pm0.0015$} \\
\bottomrule
\end{tabular}%
}
\begin{flushleft}
\scriptsize \textbf{Note:} All results are the mean of five independent runs. Standard deviations are consistently $< 0.003$ for all reported metrics. Improvements over the strongest baseline are statistically significant ($p < 0.01$).
\end{flushleft}
\end{table*}

\paragraph{Baselines}
For performance evaluation, we compare MURAL against several state-of-the-art baseline models. We first consider LightGCN \cite{he2020lightgcnsimplifyingpoweringgraph}, a foundational benchmark that simplifies the GCN architecture by removing non-linear activations, focusing solely on neighborhood aggregation. Moving into the multimodal domain, we include MMGCN \cite{10.1145/3343031.3351034} and DualGNN \cite{wang2021dualgnn}, which capture fine-grained preferences by constructing modality-specific user-item bipartite graphs and Dual graph architectures for message passing. Further, we evaluate models that incorporate latent structure discovery and graph refinement. LATTICE \cite{Zhang_2021} mines item-item correlations directly from multimodal features, FREEDOM \cite{Zhou_2023} freezes the item-item graph while denoising interaction edges, and MMGSL \cite{sone2025mmgsl} employs a graph-structured learner to model item-item correlations and adaptively fuse embeddings. Additionally, MMSSL \cite{Wei_2023} addresses label sparsity by integrating self-supervised learning with an adversarial perturbation-based structure learning paradigm. More recent generative approaches include DiffMM \cite{jiang2024diffmmmultimodaldiffusionmodel} and DiffCL \cite{11060893}. DiffMM uses a multimodal graph diffusion process to generate modality-aware graphs and mitigate data sparsity, while DiffCL primarily uses diffusion to generate augmented embeddings for contrastive learning. Finally, we compare our work with AlignRec \cite{10.1145/3627673.3679626}, which achieves cross-modal alignment via heavy cross-attention mechanisms.

\paragraph{Implementation Details}
For a rigorous evaluation, we re-implemented the baseline models using their respective open-source repositories, ensuring that all models were tested in a unified environment. We conducted exhaustive hyperparameter sweeps for all baselines, adhering to the search spaces defined in their original publications.

All experiments utilized the Adam optimizer with a fixed batch size of 1024 and a hidden embedding dimension of 64. Our proposed framework is developed in PyTorch, utilizing Xavier initialization for all weight matrices. We use a Hierarchical Navigable Small World vector database from FAISS to perform the ANN search. 

To identify optimal configurations, we used the Adam optimizer with a learning rate of $10^{-3}$. The structural hyperparameters $K_r$ and $k_m$ were selected from the ranges $[100,1000]$ and $[5,150]$, respectively. The coefficients $\lambda_{\text{ent}}$ and $\lambda_{\text{cl}}$ were sampled from a log-uniform distribution $[10^{-5}, 1.0]$. Additional dynamics, including $\beta \in \{0.2, 0.4, 0.6, 0.8, 1.0\}$, temperature $\tau \in \{0.1, 0.5, 1.0\}$, and $\lambda_{\text{reg}} \in [10^{-5}, 1.0]$, were optimized via grid search.

\subsection{Performance Comparison}
Table~\ref{tab:main_results} presents the performance results for MURAL along with the state-of-the-art baselines. 
The result presented in Table~\ref{tab:main_results} is the arithmetic mean of five independent trials conducted with distinct random seeds to ensure reproducibility and robustness.

We observe that multimodal models, including architectures like MMGCN and DualGNN, exhibit superior performance compared to LightGCN, a graph-based collaborative filtering model, thanks to their incorporation of multimodal information.
Additionally, MURAL consistently outperforms all existing models, including the most recent multimodal graph structure learning approaches. Specifically, MURAL achieves significant improvements in terms of Recall@20 and NDCG@20, validating that our retrieval-augmented strategy and uncertainty-aware fusion effectively capture underlying user preferences even in the presence of multimodal noise. 

Our empirical analysis reinforces the findings of \cite{sone2025mmgsl} regarding the efficacy of contrastive learning. Specifically, the performance gap between contrastive models (e.g., MMSSL, DiffMM) and LATTICE suggests that modeling item-item relations alone is insufficient. Contrastive signals provide the necessary supervision to bridge the gap between raw multimodal features and behavioral embeddings. MURAL achieves superior performance over MMSSL, DiffMM, and DiffCL by explicitly enriching the graph topology with latent item-item edges, unlike computationally intensive diffusion-based models that rely on stochastic denoising, our structural approach offers a more interpretable and efficient pathway for mitigating data sparsity. 

Finally, MURAL's superiority over MMGSL and FREEDOM stems from a fundamental shift from static, heuristic-bound graph construction, such as the frozen item-item graphs relied upon by FREEDOM, to a learnable topology discovery process that uncovers latent correlations during training. By explicitly modeling aleatoric uncertainty, MURAL autonomously suppresses item-level noise that deterministic frameworks cannot mitigate, ensuring robust performance across heterogeneous datasets.

\subsection{Ablation Studies}
To further investigate the contribution of each module in MURAL, we conduct a series of ablation studies. Table~\ref{tab:master_ablation} details these results. The comparison in the AEL section reveals that while the model without edge augmentation (no AEL) underperforms significantly, our full MLP-based approach consistently outpaces the static cosine similarity baseline. This gap confirms that the learned topology effectively integrates behavioral signals that standard content-based heuristics overlook, providing a more robust structural foundation for message passing.

\begin{table*}[t]
\centering
\caption{Ablation Study. We evaluate the contribution of Adaptive Edge Learning (AEL), Uncertainty-Aware Fusion (UAF), and loss components across three benchmarks.}
\label{tab:master_ablation}
\small
\setlength{\tabcolsep}{5pt}
\resizebox{\textwidth}{!}{%
\begin{tabular}{ll|cc|cc|cc}
\toprule
& \textbf{Dataset} & \multicolumn{2}{c|}{\textbf{TikTok}} & \multicolumn{2}{c|}{\textbf{Amazon-Baby}} & \multicolumn{2}{c}{\textbf{Amazon-Sports}} \\
& Metric & R@20 & N@20 & R@20 & N@20 & R@20 & N@20 \\
\midrule
\textbf{Full Model} & \textbf{MURAL (AEL-UAF)} & \textbf{0.1221} & \textbf{0.0541} & \textbf{0.1068} & \textbf{0.0502} & \textbf{0.1173} & \textbf{0.0543} \\
\midrule
\multirow{2}{*}{\textit{AEL Analysis}} 
& w/o AEL & 0.0778 & 0.0335 & 0.0715 & 0.0293 & 0.0825 & 0.0353 \\
& Static AEL (Cosine) & 0.1180 & 0.0524 & 0.1004 & 0.0417 & 0.1131 & 0.0493 \\
\midrule
\textit{Fusion Analysis} 
& Simple Averaging & 0.1153 & 0.0512 & 0.0975 & 0.0428 & 0.1043 & 0.0468 \\
\midrule
\multirow{3}{*}{\textit{Contrastive Loss}} 
& w/o $\mathcal{L}_{\mathrm{MM}}$ & 0.1181 & 0.0527 & 0.1005 & 0.0444 & 0.1116 & 0.0518 \\
& w/o $\mathcal{L}_{\mathrm{BM}}$ & 0.1107 & 0.0487 & 0.0929 & 0.0410 & 0.1039 & 0.0486 \\
& w/o $\mathcal{L}_{\mathrm{cl}}$ (All) & 0.1059 & 0.0459 & 0.0862 & 0.0395 & 0.0970 & 0.0459 \\
\midrule
\multirow{2}{*}{\textit{UAF Dynamics}} 
& w/o $\sigma$ (Attention only) & 0.1165 & 0.0517 & 0.0980 & 0.0438 & 0.1093 & 0.0511 \\
& w/o $\alpha$ (Uncertainty only) & 0.1178 & 0.0522 & 0.0991 & 0.0446 & 0.1109 & 0.0516 \\
\midrule
\multirow{3}{*}{\textit{Noise Regularization}} 
& w/o $\mathcal{L}_{\mathrm{ent}}$ (Structural Only) & 0.1140 & 0.0505 & 0.0947 & 0.0436 & 0.1071 & 0.0499 \\
& w/o $\sigma$ (Feature Only) & 0.1165 & 0.0517 & 0.0980 & 0.0438 & 0.1093 & 0.0511 \\
& w/o $\mathcal{L}_{\mathrm{ent}}$ \& w/o $\sigma$ (Unregularized) & 0.1119 & 0.0489 & 0.0922 & 0.0429 & 0.1048 & 0.0516 \\
\midrule
\textit{Gradient Detachment Analysis} 
& w/o stopgrad & 0.1091 & 0.0480 & 0.0969 & 0.0456 & 0.1055 & 0.0494 \\
\bottomrule
\end{tabular}%
}
\end{table*}

\begin{figure}[t]
    \centering
    \begin{tikzpicture}
        \begin{axis}[
            xlabel={Epoch},
            ylabel={Weight Dominance},
            grid=major,
            width=\columnwidth,
            height=4.5cm, 
            legend pos=south east,
            legend style={font=\scriptsize, cells={anchor=west}},
            xticklabel style={font=\scriptsize},
            yticklabel style={font=\scriptsize},
            ylabel style={font=\small},
            xlabel style={font=\small}
        ]
            \addplot[blue, thick] table [x=Step, y=Value, col sep=comma] {ablation_fusionfull_textdominance.csv};
            \addlegendentry{Textual}

            \addplot[red, thick] table [x=Step, y=Value, col sep=comma]{ablation_fullfusion_audiodominance.csv};
            \addlegendentry{Acoustic}

            \addplot[orange!90!black, thick] table [x=Step, y=Value, col sep=comma]{ablation_fusionfull_videodominance.csv};
            \addlegendentry{Visual}
        \end{axis}
    \end{tikzpicture}
    \caption{Evolution of modality dominance during training (TikTok). The y-axis represents the proportion of items for which a given modality holds the maximum fusion weight. UAF autonomously prioritizes high-density visual signals while attenuating non-discriminative acoustic and textual channels.}
    \label{fig:modality_dominance}
    \Description{A line graph showing the evolution of modality dominance over training epochs. The x-axis is labeled Epoch and the y-axis is labeled Weight Dominance (Proportion of Items). Three lines represent Visual, Textual, and Acoustic modalities. At the start of training, the initial dominance is random. As training progresses, the Visual line rises sharply to become the dominant modality for the vast majority of items. In contrast, the Textual and Acoustic lines decline steadily and stabilize at low values, illustrating the model's learned prioritization of visual signals.}
\end{figure}
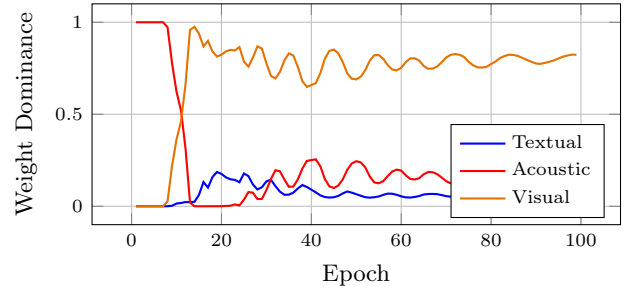

Fig.~\ref{fig:modality_dominance} illustrates the dynamic reconfiguration of modality contributions via the UAF module. We measure the dominance ratio, defined as the proportion of items in the dataset for which a specific modality receives the highest combined attention and confidence score. We observe that the model adaptively prioritizes modalities based on their discriminative power. As training converges, the visual modality emerges as the primary source of information, while textual and acoustic signals are significantly attenuated. This behavior provides empirical justification for adaptive weighting over static aggregation: 
\begin{itemize}
    \item \textbf{Information Density}: In the short-video domain, visual features exhibit the highest semantic density relative to user preference. MURAL autonomously identifies this hierarchy, capturing the platform’s primary signal without manual heuristic tuning.
    \item \textbf{Asymmetric Noise Mitigation}: Acoustic features in this context often represent non-discriminative background signals (e.g., trending audio), while textual descriptions are frequently sparse. UAF treats these as lower-reliability channels, suppressing their influence to prevent the "dilution" of high-fidelity visual representations.
    \item \textbf{Dynamic Regularization}: Unlike mean-pooling, which treats modality-specific noise and signal equally, the UAF module ensures the final representation $\tilde{h}_i$ is dominated by the most reliable latent features, effectively acting as a learned structural filter.
\end{itemize}

\begin{figure}[t]
    \centering
    \begin{tikzpicture}
        \begin{axis}[
            xlabel={Epoch},
            ylabel={Visual Uncertainty}, 
            grid=major,
            width=\columnwidth,
            height=4.5cm, 
            legend pos=south west,
            legend style={font=\scriptsize, cells={anchor=west}}, 
            xticklabel style={font=\scriptsize},
            yticklabel style={font=\scriptsize},
            ylabel style={font=\small}, 
            xlabel style={font=\small}
        ]
            \addplot[orange!90!black, thick] table [x=Step, y=Value, col sep=comma] {ablation_fusionfull_wnoise.csv};
            \addlegendentry{With Noise}

            \addplot[blue!70!black, thick] table [x=Step, y=Value, col sep=comma]{ablation_fusionfull_wonoise.csv};
            \addlegendentry{Without Noise}
        \end{axis}
    \end{tikzpicture}
    \caption{Impact of synthetic noise on Visual Uncertainty ($\sigma^{(v)}$). In the presence of noise, the UAF module autonomously attenuates the visual signal by increasing uncertainty. For clean data, it converges toward high confidence (lower $\sigma^{(v)}$).}
    \label{fig:uncertainty}
    \Description{A plot displaying how visual uncertainty evolves with and without synthetic noise. Uncertainty increases rapidly when noise is added.}
\end{figure}
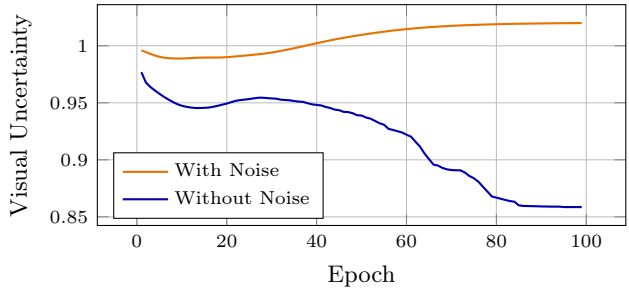

To verify that the uncertainty parameter $\sigma$ effectively captures modality-level noise, we conducted a synthetic noise-injection experiment. Fig.~\ref{fig:uncertainty} compares the evolution of the mean learned uncertainty ($\mathbb{E}[\sigma_{img}]$) for a standard "clean" training run versus a "noisy" run where visual features for 20\% of items were corrupted with Gaussian noise. In the clean setup, the mean uncertainty steadily decreases as training progresses, indicating that the model is successfully aligning the visual features with behavioral signals and gaining confidence in the modality's predictive power. Conversely, in the noisy setup, the model progressively increases the uncertainty values for the corrupted features. This divergence demonstrates that the UAF module can calibrate aleatoric uncertainty. By increasing $\sigma$, the model mathematically suppresses the contribution of noisy features in the final fused embedding $\tilde{h}_i$, thereby shielding the recommendation performance from data corruption. This adaptive mechanism provides a significant advantage over static fusion methods, which lack the ability to modulate their reliance on features based on observed reliability.

This is further corroborated by the results shown in the fusion analysis section of Table \ref{tab:master_ablation}, where we compare two variations of the model, namely, with UAF (the full model) and with simple averaging. As the results demonstrate, UAF is superior due to customized attention and uncertainty score for each item. Specifically, the ablation study confirms that $\alpha$ and $\sigma$ control different mechanisms within the fusion process and are not coupled, allowing the model to independently modulate relevance and reliability.

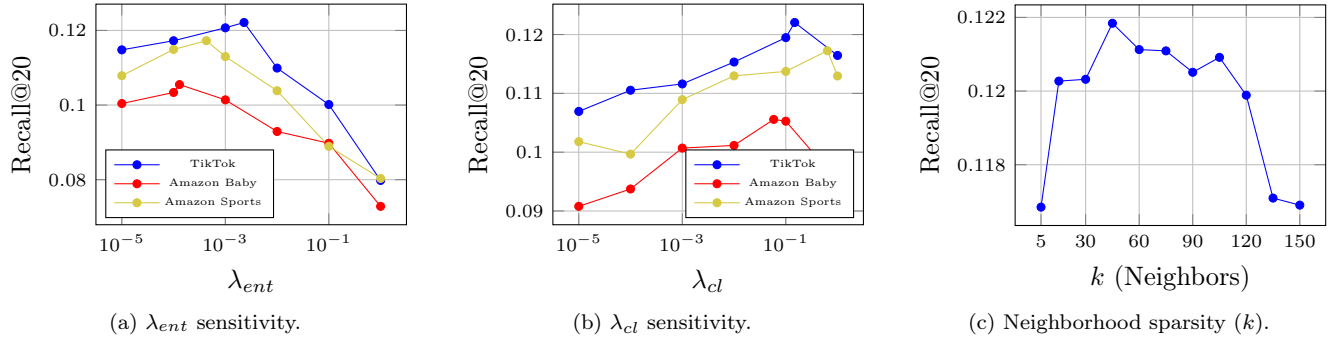
\begin{figure*}[ht]
    \centering
    \subfloat[$\lambda_{ent}$ sensitivity.\label{fig:abl_lambdaent}]{%
    \begin{minipage}[b]{0.32\textwidth}
        \centering
        \begin{tikzpicture}
            \begin{axis}[
                xlabel={$\lambda_{ent}$},
                ylabel={Recall@20},
                xmode=log,
                grid=major,
                scaled y ticks=false,
                yticklabel style={font=\scriptsize, /pgf/number format/fixed, /pgf/number format/precision=4},
                xticklabel style={font=\scriptsize},
                width=\linewidth,
                height=4.5cm,
                legend pos=south west,
                legend style={font=\fontsize{5}{6}\selectfont, nodes={scale=0.8}}
            ]
                \addplot[blue, mark=*, mark size=1.5pt] table [x=lambda_entropy, y={recall@20}, col sep=comma] {ablation_lambda_ent.csv};
                \addlegendentry{TikTok}

                \addplot[red, mark=*, mark size=1.5pt] table [x=lambda_entropy, y={recall@20}, col sep=comma] {ablation_lambda_ent_baby.csv};
                \addlegendentry{Amazon Baby}

                \addplot[yellow!80!black, mark=*, mark size=1.5pt] table [x=lambda_entropy, y={recall@20}, col sep=comma] {ablation_lambda_ent_sport.csv};
                \addlegendentry{Amazon Sports}
            \end{axis}
        \end{tikzpicture}
    \end{minipage}%
    }
    \hfill
    \subfloat[$\lambda_{cl}$ sensitivity.\label{fig:abl_lambdacl}]{%
    \begin{minipage}[b]{0.32\textwidth}
        \centering
        \begin{tikzpicture}
            \begin{axis}[
                xlabel={$\lambda_{cl}$},
                ylabel={Recall@20},
                xmode=log,
                grid=major,
                scaled y ticks=false,
                yticklabel style={font=\scriptsize, /pgf/number format/fixed, /pgf/number format/precision=4},
                xticklabel style={font=\scriptsize},
                width=\linewidth,
                height=4.5cm,
                legend pos=south east,
                legend style={font=\fontsize{5}{6}\selectfont, nodes={scale=0.8}}
            ]
                \addplot[blue, mark=*, mark size=1.5pt] table [x=lambda_cl, y={recall@20}, col sep=comma] {ablation_lambda_cl.csv};
                \addlegendentry{TikTok}

                \addplot[red, mark=*, mark size=1.5pt] table [x=lambda_cl, y={recall@20}, col sep=comma] {ablation_lambda_cl_baby.csv};
                \addlegendentry{Amazon Baby}

                \addplot[yellow!80!black, mark=*, mark size=1.5pt] table [x=lambda_cl, y={recall@20}, col sep=comma] {ablation_lambda_cl_sport.csv};
                \addlegendentry{Amazon Sports}
            \end{axis}
        \end{tikzpicture}
    \end{minipage}%
    }
    \hfill
    \subfloat[Neighborhood sparsity ($k$).\label{fig:abl_k}]{%
    \begin{minipage}[b]{0.32\textwidth}
        \centering
        \begin{tikzpicture}
            \begin{axis}[
                xlabel={$k$ (Neighbors)},
                ylabel={Recall@20},
                grid=major,
                xtick={5, 30, 60, 90, 120, 150}, 
                scaled y ticks=false,
                yticklabel style={font=\scriptsize, /pgf/number format/fixed, /pgf/number format/precision=4},
                xticklabel style={font=\scriptsize},
                width=\linewidth,
                height=4.5cm,
            ]
                \addplot[blue, mark=*, mark size=1.5pt] table [x=k, y=recall@20, col sep=comma] {ablation_k.csv};
            \end{axis}
        \end{tikzpicture}
    \end{minipage}%
    }

    \caption{Hyperparameter sensitivity analysis for $\lambda_{ent}$, $\lambda_{cl}$, and $k$. The graphs illustrate the trade-offs between model alignment, sparsity, and graph enrichment.}
    \Description{Three subplots showing hyperparameter sensitivity. Subplot a shows $\mathcal{L}_{\mathrm{ent}}$ peaking at different points for different datasets. Similarly, subplot b shows different $\mathcal{L}_{\mathrm{cl}}$ values resulting in peaks in different datasets. Subplot c shows k peaking between 30 and 120.}
    \label{fig:abl_all}
\end{figure*}

Further, we conduct an ablation study to quantify the individual contributions of each component within our joint objective function, as detailed in Table \ref{tab:master_ablation}. The contrastive loss ($\mathcal{L}_{\mathrm{cl}}$) serves as a critical semantic anchor, facilitating the alignment between behavioral trajectories and multimodal content.

Moreover, the results demonstrate that the entropy loss ($\mathcal{L}_{\mathrm{ent}}$) is essential for maintaining topological health. By maximizing neighborhood entropy, this term prevents the Adaptive Edge Learner (AEL) from collapsing into one-hot connections that would otherwise restrict the receptive field of the GNN. This encourages a more diverse and robust aggregation of item neighborhoods. Notably, our joint ablation of these regularization mechanisms confirms that they target distinct forms of corruption: $\mathcal{L}_{\mathrm{ent}}$ primarily suppresses structural noise within the graph topology, whereas $\sigma$ explicitly suppresses content-level feature noise. Acting as complementary filters, they ensure that both clean node representations and diverse semantic pathways consistently contribute to the final recommendation.

Finally, we empirically evaluate the necessity of the gradient detachment applied to $h_i$ (Eq.~\ref{eq:aligned}). As shown in the final row of Table~\ref{tab:master_ablation}, removing the $\operatorname{stopgrad}$ operation leads to a noticeable performance degradation across all datasets (e.g., Recall@20 drops from $0.1221$ to $0.1091$ on TikTok). This drop confirms that isolating the behavioral gradients is crucial for preventing the modality alignment process from leaking back into and distorting the underlying collaborative topology.

\subsection{Hyperparameter Effects}
We investigate the sensitivity of recommendation performance to the loss weight variations of $\lambda_{\mathrm{cl}}$ and $\lambda_{\mathrm{ent}}$ (Figs.~\ref{fig:abl_lambdacl} and~\ref{fig:abl_lambdaent}). Our results indicate that while both parameters are evaluated within the same numerical range, their optimal configurations are highly dataset-dependent. Specifically, the contrastive weight $\lambda_{\mathrm{cl}}$ must balance semantic alignment with the primary ranking objective, while the entropy weight $\lambda_{\mathrm{ent}}$ regulates the sparsity of the learned graph. These findings underscore that careful calibration of these hyperparameters is essential for maximizing the model's robustness and achieving superior accuracy across diverse multimodal environments.

Sensitivity analysis on neighborhood sparsity (Fig.~\ref{fig:abl_k}, TikTok) shows that increasing $k$ initially improves Recall@20 by reducing structural distance and facilitating efficient message passing. However, performance degrades beyond an optimal threshold as weakly correlated item pairs introduce semantic noise, distorting user-item relationships. Notably, performance remains stable around the peak, suggesting the UAF module acts as a learned regularizer. By assigning personalized uncertainty scores, the model dynamically weighs modality signals and mitigates sensitivity to graph density. This highlights the importance of balancing graph enrichment with structural signal preservation.

\section{Conclusion and Future Work}
In this work, we introduced MURAL, a unified framework that overcomes the fundamental bottlenecks of structural rigidity and semantic fragility in multimodal recommendation. By transitioning from static, pre-defined similarity heuristics to dynamic topology discovery, MURAL provides a robust and scalable architecture capable of navigating the inherent noise and complexity of modern heterogeneous data. Our results demonstrate that the challenges of "graph mirroring" and cross-modal noise are not insurmountable obstacles, but rather symptoms of non-adaptive modeling. Through the Adaptive Edge Learner (AEL), we proved that treating graph construction as a differentiable, retrieval-augmented task allows for the discovery of latent semantic correlations that evolve in tandem with user behavior. Simultaneously, the Uncertainty-Aware Fusion (UAF) module provides a mathematically grounded defense against modality noise. By explicitly modeling aleatoric uncertainty ($\sigma_i^{(m)}$), our framework moves beyond deterministic fusion to prioritize high-fidelity signals, ensuring that recommendations are driven by the most reliable modality-specific insights. Extensive empirical evaluations confirm that this unified approach to topology learning and noise-robustness consistently outperforms current state-of-the-art structural and generative models. Ultimately, MURAL establishes a new benchmark for interpretable and efficient multimodal learning, offering a resilient path forward for the next generation of content-aware recommendation systems.

While MURAL significantly improves recommendation accuracy and robustness, several avenues for future research remain. First, extending the uncertainty-aware mechanism to the user side-modeling user-specific variance in modality preference---could further personalize the fusion process. Second, exploring the temporal dynamics of the adaptive graph (i.e., how semantic neighbors shift over longer training horizons) may yield deeper insights into evolving item trends. Ultimately, we believe the principles of adaptive structure learning and reliability-aware fusion established in this work provide a robust foundation for the next generation of scalable, multimodal GNNs.

\section*{Acknowledgment}
The authors declare that they have no financial or non-financial interests that are directly or indirectly related to the work submitted for publication; that they have not received any funding, financial support, or sponsorship from any organization or agency for the preparation of this work; and that during the preparation of this work, they used generative AI technologies with extreme caution to improve language and readability, after which they reviewed and edited the content as needed and take full responsibility for the publication's content.

\bibliography{references}

\end{document}